\documentclass[aps,showpacs,preprintnumbers,amsmath,amssymb]{revtex4}

\UseRawInputEncoding
\usepackage{graphicx}
\usepackage{epstopdf}
\usepackage{color}
\usepackage{subfig}
\usepackage{aasmacros}
\usepackage{accents}
\usepackage{natbib}

\begin{document}
    \captionsetup{justification=raggedright,singlelinecheck=false}

    \baselineskip=0.8cm
    \title{\bf Constraining a disformal Schwarzschild black hole in DHOST theories with the orbit of the S2 star}

    \author{Zelin Zhang$^{1}$\footnote{zzl@hunnu.edu.cn},
    Songbai Chen$^{1,2}$\footnote{Corresponding author: csb3752@hunnu.edu.cn},
    Jiliang Jing$^{1,2}$ \footnote{jljing@hunnu.edu.cn}}
    \affiliation{$^1$Department of Physics, Institute of Interdisciplinary Studies, Key Laboratory of Low Dimensional Quantum Structures
    and Quantum Control of Ministry of Education, Synergetic Innovation Center for Quantum Effects and Applications, Hunan
    Normal University,  Changsha, Hunan 410081, People's Republic of China
    \\
    $ ^2$Center for Gravitation and Cosmology, College of Physical Science and Technology, Yangzhou University, Yangzhou 225009, People's Republic of China}

    \begin{abstract}
    \baselineskip=0.6 cm
    \begin{center}
    {\bf Abstract}
    \end{center}
        With the observed data of the S2 orbit around the black hole Sgr A$^*$ and the Markov Chain Monte Carlo method, we make a constraint on parameters of a disformal Schwarzschild black hole in quadratic degenerate higher-order scalar-tensor (DHOST) theories. This black hole belongs to a class of non-stealth solutions and owns an extra disformal parameter described the deviation from general relativity. Our results show that the best fit value of the disformal parameter is positive. However, in the range of $1\sigma$, we also find that general relativity remains to be consistent with the observation of the S2 orbit.

    \end{abstract}

    \maketitle
    \newpage

    \section{Introduction}

    Einstein's general relativity (GR) is widely believed to be the most successful theory of gravity, which has successfully passed a series of tests from observations and experiments \cite{2014LRR....17....4W,2016PhRvL.116f1102A,2019Sci...365..664D,2019ApJ...875L...1E,2022ApJ...930L..12E}.
    However, it is undoubtedly that GR faces some theoretical and observational difficulties including singularity problem and dark matter. Moreover,the current observations cannot completely exclude the possibility of other alternative theories of gravity with deviation from GR.
    Therefore, it is necessary to probe signals from alternative theories of gravity and to further test such kind of gravities.

    One of important alternative theories is the so-called scalar-tensor theory where there exists one scalar degree of freedom besides the gravitational field. The degenerate higher-order scalar-tensor (DHOST) theories belong to such kind of scalar-tensor theories      \cite{2016JCAP...02..034L,2016JCAP...07..016L,2016PhRvD..93l4005B,2016JCAP...04..044C}, which contain higher order derivatives of scalar field and meet a certain set of degeneracy conditions.  DHOST theories can be also looked as a kind of extensions of both Horndeski theories \cite{1974IJTP...10..363H} and beyond Horndeski theories \cite{2014ffp..confE.102V}. Although there exists higher-order equations of motion, DHOST theories can aviod the Ostrogradsky ghost due to satisfying degeneracy conditions. For the DHOST theories, there exist two distinct types of black hole solutions \cite{2019PhRvD..99f4042B,2019PhRvD..99f4040M,2019PhRvD.100d4053M,2019PhRvD.100h4020C,2020JCAP...06..034T,2020JCAP...02..023B,2020JCAP...06..049B,2020CQGra..37c5007V,2023arXiv230912229B}. One of them is a type of stealth solutions in which the extra scalar field is absented in the spacetime metric and the solution owns the same metric as that in GR. The other is the so-called non-stealth solutions and their metrics differ from that in Einstein's theory because the parameters of scalar field appear in the metrics \cite{2021arXiv210311788C,2020EPJC...80.1180L,2021CQGra..38g5026C,2021PhRvD.103l4035A,2022SCPMA..6550411Z,2023PASJ...75S.217T}. Recently, a rotating black hole solution in quadratic DHOST theories \cite{2021JHEP...01..018A,2020JCAP...11..001B} is obtained by the disformal and conformal transformations \cite{2016PhRvD..93l4005B}, which belongs to non-stealth solutions because of existence of extra disformal parameter arising from the scalar field.
    The deformation parameter could give rise to some new observational effects differed from those in the Kerr case. It is found that the deformation parameter heavily affect the black hole shadow so there are some eyebrowlike shadows with self-similar fractal structures \cite{2020EPJC...80.1180L}.
    The effects of the deformation parameter on the orbital precession are analyzed for the post-Newtonian motion of stars orbiting disformal black holes by using the osculating orbit method \cite{2021PhRvD.103l4035A}. Moreover, the disformal parameter is found to appear in the frequencies of quasi-periodic oscillations and the corresponding constraint is performed by using  the  observation data of GRO J1655-40 \cite{2021arXiv210311788C}.
    It is  expected to further probe the spacetime geometry of a disformal rotating DHOST black hole  and to measure its disformal parameter by future tests with the Square Kilometer Array (SKA) \cite{2023PASJ...75S.217T} and LISA \cite{2024arXiv240316192B}.

    It is well known that there is a supermassive black hole Sgr A* inhabited the center of Milky Way galaxy and the orbits of S-stars around  Sgr A* have been detected through the last three decades \cite{2002Natur.419..694S,2003ApJ...586L.127G,2005ApJ...628..246E,2008ApJ...689.1044G,2009ApJ...692.1075G,2009A&A...502...91S,2012Sci...338...84M,2016ApJ...830...17B,2017ApJ...837...30G,2018A&A...615L..15G,2019Sci...365..664D,2020A&A...636L...5G}. The S2 star is the most notable in the cluster around  Sgr A*, whose orbit is characterized by a short period of $\sim 16$ years, a semi-major axis of $\sim 970$ AU and an high eccentricity of $\sim 0.88$. Moreover, the speed of the S2 star at the pericentre is about $7700 km/s$, which is close to $2.6\%$ of the speed of light \cite{2018A&A...615L..15G,2020A&A...636L...5G}. These peculiar orbital features of the S2 star enable to  detect the special and general relativistic effects, which ensures that S2 serves as a direct probe for space-time metric and further provides a great opportunity to test theories of gravity in the strong regime. With the observation data of the S2 star including the gravitational redshift\cite{2018A&A...615L..15G}, the relativistic redshift\cite{2019Sci...365..664D}, and  the precession \cite{2020A&A...636L...5G}, a lot of works have been already carried out to examine the no-hair theorem \cite{2023EPJC...83..311L,2023PhRvD.107d4038C} , to probe dark matter \cite{2020arXiv200606219T,2020A&A...641A..34B,2022A&A...660A..13H,2022PhRvD.106j3024Y,2023arXiv230309284S,2023MNRAS.524.1075F} and to test theories of gravity \cite{2021Univ....7..407B,2022Univ....8..137D,2022JCAP...09..008Y,2023JCAP...08..039F,2022EPJC...82..633J,2022EPJC...82.1018J,2022MNRAS.510.4757D,2023arXiv230710804S,2023JCAP...03..056J,2023MNRAS.519.1981D,2023ApJ...944..149A,2023PhRvD.107h4043Y}.

    In this paper, we will focus on the disformal black hole in DHOST theories in the previous discussion.  As in \cite{2015ApJ...809..127Z,2017A&A...608A..60G}, since the observational effects arising from the rotation of the central black hole on the S2 star's orbit are expected to be very small, we here also ignore the effects of the angular momentum of the spacetime and consider only the disformal Schwarzschild spacetime.

    The paper is organized as follows: In Sec.II,  we present a very brief introduction of the disformal Schwarzschild black hole spacetime and model the orbital motion of S2 in this spacetime. In Sec.III, we carry out the analysis of the MCMC implemented by emcee and fit the black hole parameters with the observational data from the S2 star. Finally, we end the paper with a summary.

    \section{Modelling the orbital motion of S2 around a disformed Schwarzschild black hole in DHOST theories}

        We firstly review in brief the disformal Schwarzschild black hole DHOST theories. This black hole belongs to a class of non-stealth solutions and owns an extra disformal parameter described the deviation from GR. The general action in the quadratic DHOST theories can be expressed as \cite{2016JCAP...02..034L,2016JCAP...07..016L,2016PhRvD..93l4005B,2016JCAP...04..044C}
        \begin{eqnarray}
        S=\int d^{4} x \sqrt{-g}\left(P(X, \phi)+Q(X, \phi) \square \phi+F(X, \phi) R+\sum_{i=1}^{5} A_{i}(X, \phi) L_{i}\right),\label{eqn:S}
        \end{eqnarray}
        with
        \begin{eqnarray}
        L_{1} &\equiv& \phi_{\mu \nu} \phi^{\mu \nu}, \quad L_{2} \equiv(\square \phi)^{2}, \quad L_{3} \equiv \phi^{\mu} \phi_{\mu \nu} \phi^{\nu} \square \phi, \nonumber\\
        L_{4} &\equiv& \phi^{\mu} \phi_{\mu \nu} \phi^{\nu \rho} \phi_{\rho}, \quad L_{5} \equiv\left(\phi^{\mu} \phi_{\mu \nu} \phi^{\nu}\right)^{2},
        \end{eqnarray}
        where $R$ is the usual Ricci scalar and $\phi$ is the scalar field.  $X \equiv \phi_{\mu} \phi^{\mu}$ is the kinetic term of the scalar field and  $\phi_{\mu} \equiv \nabla _\mu \phi$ is the corresponding covariant derivative. The quantities $P$, $Q$, $F$ and $A_i$ are functions of $\phi$ and $X$.
        In the quadratic DHOST theory, the functions $P$  and $Q$ are totally free, but $F$ and $A_i$  must satisfy certain degeneracy conditions, which ensures that there is only an extra scalar degree of freedom besides the usual
        tensor modes of gravity. The degeneracy conditions for quadratic DHOST Ia theory is given in \cite{2016PhRvD..93l4005B}.
        It is well known that a new solution within DHOST Ia theory can be obtained from the known ``seed" solution by performing a disformal transformation.
        In general, the disformal transformation between the ``disformed" metric $g_{\mu \nu}$ and the original ``seed" metric $\tilde{g}_{\mu \nu}$ can be expressed as \cite{2016PhRvD..93l4005B}
        \begin{equation}
        g_{\mu \nu}=A(X, \phi) \tilde{g}_{\mu \nu}-B(X, \phi) \phi_{\mu} \phi_{\nu},\label{eq:metric1}
        \end{equation}
        where $A(X,\phi)$ is a conformal factor and $B(X,\phi)$ is a disformal one.
        To obtain a new solution, the functions $A$ and $B$ must meet the conditions to ensure that the ``disformed" metric $g_{\mu \nu}$ does not
        degenerate into the original ``seed" metric $\tilde{g}_{\mu \nu}$.
        Starting from the usual Schwarzschild metric, one can obtain the disformal Schwarzschild solution by making use of the transformation with $A(X, \phi)=1$ and $B(X, \phi)=B_0$ ($B_0$ is a constant) \cite{2021JHEP...01..018A,2020JCAP...11..001B}
        \begin{equation}
        ds^{2}=(1-\frac{2M}{r}-\alpha)dt^{2}-\frac{r^2+ 2 (\alpha -1) Mr}{(r-2 M)^2} dr^{2}-\frac{2 \sqrt{2 M r} \alpha} {r - 2 M} dt dr-r^2 d\theta^{2}-r^2 \sin^2 \theta d\phi^2,
        \label{eq:metric}
        \end{equation}
       and the scalar field has a form depended on only the coordinates $t$ and $r$ \cite{2021JHEP...01..018A,2020JCAP...11..001B}
        \begin{eqnarray}
        &&\phi(t, r)=-m t+S_{r}(r), \quad\quad\quad S_{r}= -m\int  \frac{\sqrt{2Mr}}{r-2M} d r.\label{eqn:scalar}
        \end{eqnarray}
        Interestingly, this particular form of the scalar field  can avoid the pathological behavior of the disformal metric (\ref{eq:metric}) at spatial infinite \cite{2021JHEP...01..018A,2020JCAP...11..001B}. The parameter $\alpha$ is connect with the scalar field's rest mass $m$ by the relation $\alpha=-B_{0}m^2$. The choice of $A(X, \varphi)=1$ is made to prevent a globally irrelevant constant conformal factor in the metric. Comparing with the usual Schwarzschild case, the disformal Schwarzschild metric (\ref{eq:metric}) owns an extra disformal parameter $\alpha$, which decribes the deviation from GR. Moreover,  the presence of the $drdt$ term yields some properties differed from the usual Schwarzschild spacetime.

        The trajectory of a free massive test particle within the disformal Schwarzschild space-time is governed by its timelike geodesic equation with a form
        \begin{equation}\label{eq:geo}
            \frac{d^2x^\mu}{d\tau^2}+\Gamma^\mu{}_{\nu \rho}\frac{dx^\nu}{d\tau}\frac{dx^\rho}{d \tau}=0,
        \end{equation}
        where $\tau$ denotes the proper time and $\Gamma^\mu{}_{\nu \rho}$ are the Christoffel symbols for the disformal Schwarzschild spacetime.
        As in \cite{2022JCAP...09..008Y,2022Univ....8..137D,2023JCAP...08..039F}, we limit the test particle's motion in the equatorial plane ($\theta = \pi/2$, $\dot{\theta} =\ddot{\theta} = 0$) and then the geodesic equation (\ref{eq:geo}) can be simplified as
        \begin{eqnarray}
                \dot{t} &=& \dfrac{E+g_{tr}\dot{r}}{g_{tt}},\label{eqn:motiont} \\
                \ddot{r} &=& \frac{g_{tt} (g_{tt,r} \dot{t}^2 - g_{rr,r} \dot{r}^2   + g_{\phi \phi,r} \dot{\phi}^2 )+2g_{tr} \dot{r}(g_{tt,r} \dot{t} +g_{tr,r} \dot{r} )}{2g_{rr} g_{tt}-2 g_{tr}^2},\label{eqn:motionr} \\
                \dot{\phi} &=&  -\dfrac{L}{g_{\phi \phi}},\label{eqn:motionphi}
            \end{eqnarray}
        where $E$ and $L$ represent the conserved energy and the angular momentum of the particle, respectively. From Eq.(\ref{eqn:motiont}), the quantities $\dot{t}$ depends on $\dot{r}$ due to the presence of $g_{tr}$. Consequently, the motion of the timelike particle exhibits distinct characteristics differed in the case of a Schwarzschild black hole in GR. From the normalization condition $g_{\mu\nu}\dot{x}^{\mu}\dot{x}^{\nu} = 1$, we have
        \begin{equation}
            \left(g_{tr}^2-g_{tt}g_{rr}\right)\dot{r}^2 = E^2 - V_{eff},
        \end{equation}
        where the effective potential is
        \begin{equation}
            V_{eff} = g_{tt} \left(1-\frac{L^2}{g_{\phi \phi}}\right).
        \end{equation}
        For a timelike test paraticle moving around a black hole, one can determine its periastron $r_p$ and apastron $r_a$ by $\dot{r} = 0$. Then, we can get its two conserved quantities $E$ and $L$ by solving equations $E^2 = V_{eff}(r_p) = V_{eff}(r_a)$. With these two values, the equations (\ref{eqn:motiont})-(\ref{eqn:motionphi}) can be numerically integrated using the initial coordinates $(t_0, r_0, \dot{r}_0, \phi_0)$. Here, we use the DifferentialEquations in Julia \cite{rackauckas2017differentialequations} to solve dynamical equations (\ref{eqn:motiont})-(\ref{eqn:motionphi}) both forward and backward in time by setting the periastron as the starting point ($t_0 = 2018.35+\Delta_t$, $r_0 = r_p$, $\dot{r}_0 = 0$, $\phi_0 = 0$).
        Then one can get the sky-projected position coordinates from integrated orbit by using classic Thiele-Innes constants \cite{2019Sci...365..664D,2020A&A...641A..34B,2022JCAP...09..008Y}. It is necessary to consider offsets and linear drifts between the gravitational center and the origin of reference frame. We introduce four additional parameters ($x_0$, $y_0$, $v_{x0}$, $v_{y0}$) to model the offsets and linear drifts \cite{2019Sci...365..664D,2022JCAP...09..008Y}. The radial velocity $RV$ of the S2 star can be obtained by $RV = c \cdot \zeta + v_{z0}$, where $v_{z0}$ is a constant velocity offset and $\zeta$ is the photon’s frequency shift. The quantity $\zeta$ describes combination of the relativistic Doppler shift and the gravitational redshift \cite{2022JCAP...09..008Y}. We also calculate orbital precession $\Delta \phi = \phi_1|_{\dot{r}=0} - \phi_0|_{\dot{r}=0}-2\pi$ in the model, and here $\phi|_{\dot{r}=0}$ represents the coordinate $\phi$ at radial turning points. Furthermore, R\"omer time delay is a significant factor that is also taken into consideration in the model as in ref. \cite{2019Sci...365..664D}. In this way, we can fit the observed apparent orbital data with the apparent orbital plane and probe the effects of disformal Schwarzschild black hole.

    \section{data and fitting}

    In this section, we  carry out the analysis of the MCMC implemented by emcee \cite{2013PASP..125..306F,2006S&C....16..239T} to obtain the constraints on the disformed Schwarzschild black hole spacetime.  Here, we utilize a dataset containing 46 sets of astrometric positions, 115 sets radial velocities of S2 \cite{2019Sci...365..664D}, and a set of relativistic orbital precession obtained  by the Gravity Collaboration \cite{2020A&A...636L...5G}.
    Our fitting model consists of six orbital parameters of S2 ($a$, $e$, $\omega$, $i$, $\Omega$, $\Delta_t$), and 9 global parameters including the Black Hole mass $M$, the distance  $R_0$ between Galactic Center and observer, the deformation parameter $\alpha$, the 2-D position and velocity of the Black Hole ($x_0$, $y_0$, $v_{x0}$, $v_{y0}$) with an offset for its radial velocity ($v_{z0}$), and an offset from the Near-Infrared Camera 2 (NIRC2) measurements ($\mathrm{offset}$).

       In our MCMC analysis, the likelihood function is given by
        \begin{equation}
            \log \mathcal{L} = \log \mathcal{L}_{\rm Pos} + \log \mathcal{L}_{\rm VR} + \log \mathcal{L}_{\rm Pre},
        \end{equation}
       where $\log\mathcal{L}_{\rm Pos}$, $\log\mathcal{L}_{\rm VR}$  and $\log\mathcal{L}_{\rm Pre}$ respectively denote the likelihood of the positional data, the radial velocities and the orbital precession. Their form can be expressed as
       \begin{equation}
            \log\mathcal{L}_{\rm Pos} = -\frac{1}{2}\sum_i\left(\frac{(x_{\rm obs}^i-x_{\rm th}^i)^2}{{(\sqrt{2}\sigma_{x,{\rm obs}}^i})^2}-\frac{(y_{\rm obs}^i-y_{\rm  th}^i)^2}{{(\sqrt{2}\sigma_{y,{\rm obs}}^i})^2}\right),
        \end{equation}
        \begin{equation}
            \log\mathcal{L}_{\rm VR} = -\frac{1}{2}\sum_i\frac{(\textrm{RV}_{\rm obs}^i-\textrm{RV}_{\rm th}^i)^2}{{(\sqrt{2}\sigma_{\textrm{RV},{\rm obs}}^i})^2},
        \end{equation}
        and
        \begin{equation}
            \log\mathcal{L}_{\rm Pre} = -\frac{1}{2}\frac{(f_{\rm SP, obs}-f_{\rm SP, th})^2}{(\sqrt{2}\sigma_{f_{\rm SP, obs}})^2}.
        \end{equation}
        Here $f_{\rm SP, th}\equiv\Delta\phi_{\rm DHOST}/\Delta\phi_{\rm GR}$. The subscript ``obs" denotes the observed data, ``th" represents the corresponding theoretical values, and $\sigma$ are statistical uncertainty for the associated quantities. The additional factor $\sqrt{2}$ in the denominators of the likelihood function is introduced to avoid double counting data. The priors for the Keplerian orbital elements of S2 are adopted to uniform probability distributions. Moreover, for the parameters related to the drift and offset of the reference frame, we assign Gaussian priors centered on the estimation in \cite{2019Sci...365..664D}. These priors are list in the table \ref{Tab:prior}.
        \begin{table}[!htbp]
            \centering
            \begin{tabular}{cccc}
                \hline
                \hline
                & \multicolumn{2}{c}{Uniform priors}& \\
                \cline{2-3}
                Parameter & Start & End  \\
                \hline
                $R(kpc)$ & $2$ & $20$  \\
                $M(10^6 M_{\odot})$ & $2$ & $20$  \\
                $a(mas)$ & $100$ & $150$  \\
                $e$ & $0.8$ & $0.95$ \\
                $i(deg)$ & $0$ & $360$ \\
                $\omega(deg)$ & $0$ & $360$  \\
                $\Omega(deg)$ & $0$ & $360$  \\
                $t_p-2018.35(yr)$ & $-1$ & $1$ \\
                $\mathrm{offset}(km \cdot s^{-1})$ & $-300$ & $300$  \\
                $\alpha$ & $-1$  & $1$  \\
                \hline
                & \multicolumn{2}{c}{Gaussian priors}& \\
                \cline{2-3}
                Parameter & $\mu$ & $\sigma$  \\
                \hline
                $x_0(mas)$ & $1.22$ & $0.32$  \\
                $y_0(mas)$ & $-0.88$ & $0.34$  \\
                $v_{x0}(mas \cdot yr^{-1})$ & $-0.077$ & $0.018$  \\
                $v_{y0}(mas \cdot yr^{-1})$ & $0.226$ & $0.019$  \\
                $v_{z0}(km \cdot s^{-1})$ & $-6.2$ & $3.7$ \\
                \hline
                \hline
            \end{tabular}
            \caption{Prior settings applied to our analysis of the S2 orbital model within the framework of DHOST theory. The Gaussian priors of ($x_0$, $y_0$, $v_{x0}$, $v_{y0}$, $v_{z0}$) centered on the estimation in \cite{2019Sci...365..664D} }\label{Tab:prior}
        \end{table}

\begin{figure}[htbp]
            \centering
            \includegraphics[width=\hsize,clip]{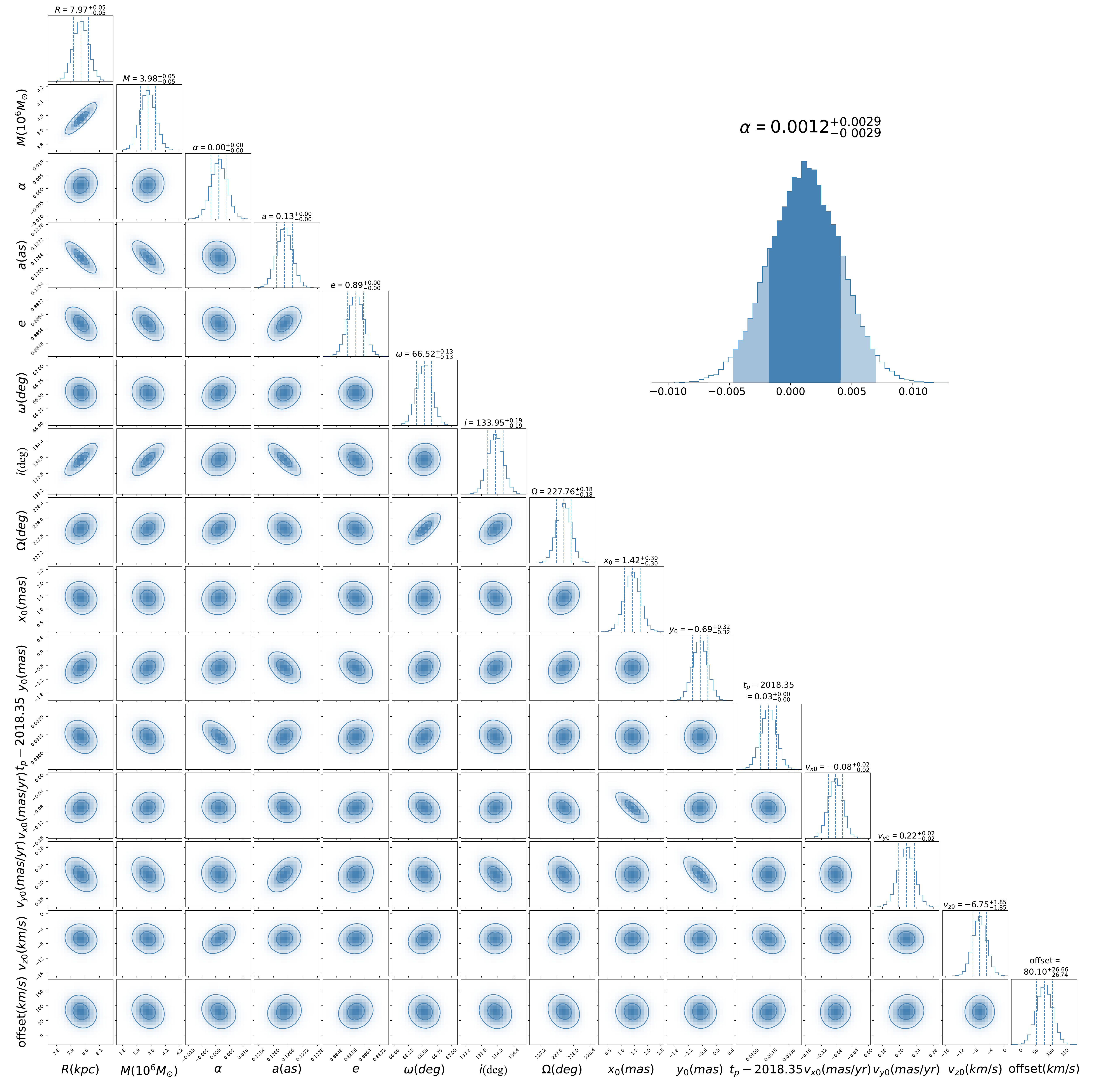}
            \caption{The corner plot \cite{corner} of the 15 parameters obtained from an MCMC sample. The Estimation value of the parameter and the range of $1-\sigma$ are marked at the top of each column.}\label{fig:corner}
        \end{figure}

        In Fig. \ref{fig:corner}, we present the full posterior distributions spanning the 15-dimensional parameter space of our model. The shaded regions show the $1$-$\sigma$ and $2$-$\sigma$ confidence levels (C.L.) of the posterior probability density distributions for the complete set of parameters, respectively. In Table \ref{Tab:results}, we list the constrained values for all 15 model parameters with their corresponding parameter ranges within the $1-\sigma$ confidence level. Our results show that in the $1$-$\sigma$ range the disformal parameter is constrained to $\alpha=0.0012^{+0.0029}_{-0.0029}$, which indicates that the constraint of $\alpha$ from the S2 orbit is more precise than that from quasi-periodic oscillations \cite{2021arXiv210311788C} where the disformal parameter is limited in $\alpha=-0.01_{-0.012}^{0.011}$ at confidence level at $68.3\%$.
        Our results show that the case of $\alpha=0$  still lies in the range of $1$-$\sigma$, which means that general relativity remains to be consistent with the observation data of the S2 orbit around the black hole Sgr A* in our galaxy. The theoretical analyses on the SKA detection \cite{2023PASJ...75S.217T} for pulsars evolving from stars like S2 implies the effects from the disformal parameter can be observable as $|\alpha| > 0.1$. With the disformal Kerr black hole metric, the study of the extreme mass-ratio inspirals (EMRIs) gravitational wave indicates that the minimal disformal parameter $\alpha$ that could be detected by LISA changes from $10^{-5}$ to $10^{-4}$ in the case of Sgr A* \cite{2024arXiv240316192B}.

        From the relationship $\alpha=-B_0 m^2$, the range of the dimensional parameter $B_0$ depends on the mass $m$ of the scalar field. The mass range of the scalar field around Sgr A* has been estimated to $10^{-20} \text{ eV} \lesssim m \lesssim 10^{-18} \text{ eV}$  from the S2 star \cite{2019MNRAS.489.4606G} and to  $10^{-18} \text{ eV} \lesssim m \lesssim 10^{-17} \text{ eV}$ from  the polarimetric images by the Event Horizon Telescope \cite{2020PhRvL.124f1102C}, respectively. Based on these estimated ranges of $m$, we can find that the value of $B_0$ lies in the range $-5.2 \times 10^6 \;\text{s}^2 \lesssim B_0<0$ for the scalar field mass with $m\sim10^{-20} \text{ eV}$, and $-17.8 \;\text{s}^2\lesssim B_0<0$  for the scalar field mass $m\sim10^{-17} \text{ eV}$. With these ranges of $B_0$, we can get the largest lower bound for the suppression scale $\Lambda$ is around $\Lambda \gtrsim 0.084 \text{ MeV}$.  The suppression scale $\Lambda$ is also estimated from other aspects \cite{2010JCAP...05..038Z,2012PhRvL.109x1102K,2013PhRvL.111p1302V,2013JCAP...11..001B,
        2014A&A...569A..90N,2014JCAP...12..012S,2014PhRvD..90j4009B,2015PhRvD..92d4036B,2015JCAP...10..051I,2018PhRvD..98f3531B,
        2022PhRvD.105b4052B,2023PhRvD.108f3031B}. The mono-jet searches performed by the CMS collaboration provide the best current constraint on the energy scale $\Lambda \gtrsim 650 \text{ GeV}$ \cite{2015PhRvD..92d4036B}. With the geodetic and frame-dragging effects of the satellite experiments within the solar system, the suppression scale $\Lambda$  is also constrained to be greater than $10^{-18} \text{ eV}$ \cite{2023PhRvD.108f3031B}.
        For the galactic center black hole Sgr A*,  the orbital behavior of nearby stars coupled with scalar fields suggests that the suppression scale $\Lambda  \sim 0.08 \text{ MeV}$ \cite{2022PhRvD.105b4052B}, and the frame-dragging precession of the star S2 yields a lower bound  $\Lambda > 10^{-5} \text{ eV}$ \cite{2023PhRvD.108f3031B}.  Thus, our constraint on the suppression scale $\Lambda$ from the star S2  motion is consistent with that obtained in \cite{2022PhRvD.105b4052B}.

        \begin{table}[htbp]
            \centering
            \begin{tabular}{ccc}
                \hline
                \hline
                Parameter & Description & MCMC Result \\
                \hline
                $R(kpc)$ & Distance to GC & $7.971_{-0.052}^{0.053}$ \\
                $M(10^6 M_{\odot})$ & Black Hole Mass & $3.978_{-0.052}^{0.053}$ \\
                $a(mas)$ & Semimajor Axis & $126.452_{-0.317}^{0.318}$ \\
                $e$ & Eccentricity & $0.886_{-0.00044}^{0.00044}$ \\
                $t_p-2018.35(yr)$ & Closest Approach & $0.031_{-0.00065}^{0.00064}$ \\
                $i(deg)$ & Inclination & $133.95_{-0.187}^{0.189}$ \\
                $\omega(deg)$ & Argument of Periapsis & $66.52_{-0.132}^{0.131}$ \\
                $\Omega(deg)$ & Angle to the Ascending Node & $227.76_{-0.176}^{0.176}$ \\
                $x_0(mas)$ & x Dynamical Center & $1.418_{-0.300}^{0.296}$ \\
                $y_0(mas)$ & y Dynamical Center & $-0.694_{-0.320}^{0.323}$ \\
                $v_{x0}(mas \cdot yr^{-1})$ & x Velocity & $-0.084_{-0.018}^{0.018}$ \\
                $v_{y0}(mas \cdot yr^{-1})$ & y Velocity & $0.217_{-0.019}^{0.019}$ \\
                $v_{z0}(km \cdot s^{-1})$ & z Velocity & $-6.753_{-1.850}^{1.850}$ \\
                NIRC2 offset$(km \cdot s^{-1})$ & NIRC2 RV Offset & $79.85_{-27.01}^{26.46}$ \\
                \hline
                $\alpha$ & Disformal Parameter  & $0.0012_{-0.00293}^{0.00288}$ \\
                \hline
                \hline
            \end{tabular}
            \caption{Estimation values for the parameters of the S2 orbital model in DHOST gravity theory, obtained through MCMC analysis.}\label{Tab:results}
        \end{table}
         \begin{figure}[htbp]
            \centering
            \includegraphics[width=8cm,clip]{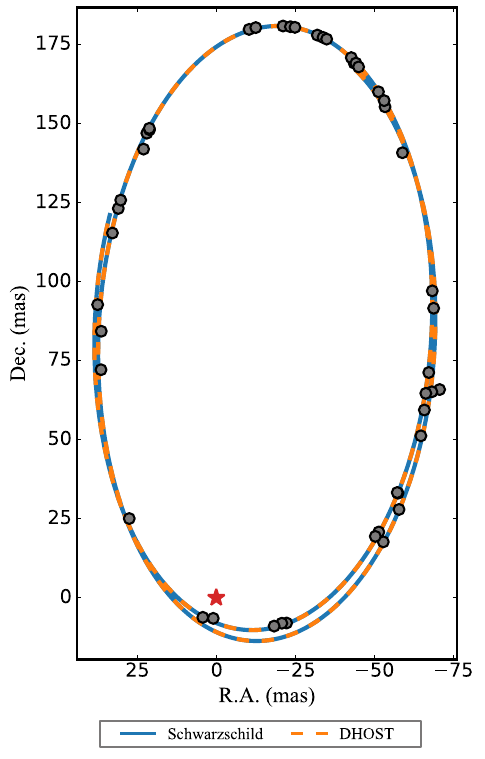}
            \caption{Astrometric observed positions (gray points) of S2 orbiting around Sgr A* (red star) and its best-fitting orbits.  The dashed and solid lines correspond to the disformal Schwarzschild black hole in DHOST theories and the usual Schwarzschild case in GR, respectively. The unit as is arcseconds and the origin of coordinates is the position of Sgr A*.}\label{fig:orb}
        \end{figure}
        \begin{figure}[htbp]
            \centering
            \includegraphics[width=8cm,clip]{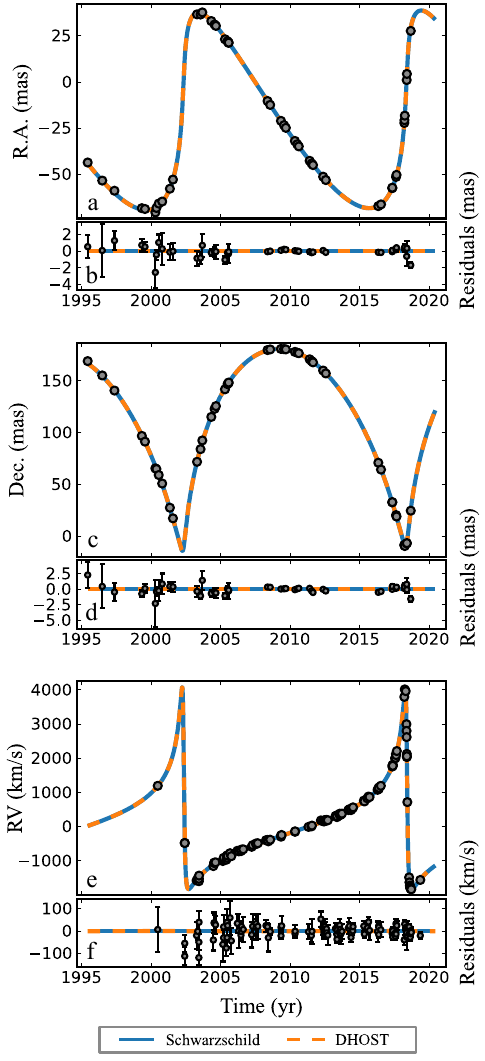}
            \caption{The data of Right Ascension (R.A.), Declination (Dec.) and RV of the S2 star (the green points) and the corresponding fitting curves. The dashed and solid lines correspond to the disformal Schwarzschild black hole in DHOST theories and the usual Schwarzschild case in GR, respectively.
             }\label{fig:xyz}
        \end{figure}

        Finally, in Figs. \ref{fig:orb} and \ref{fig:xyz}, we present the best-fitting orbit of S2 and the corresponding fitting curves of Right Ascension (R.A.), Declination (Dec.) and RV for the disformal Schwarzschild black hole in DHOST theories and the usual Schwarzschild black hole in GR.  Our results show  that  the disformal Schwarzschild black hole in DHOST theories with the best-fitting parameters can also describe the strong gravitational field of the black hole Sgr A* in our Milkyway.

    \section{Summary}

    In this paper, we focus on the effects of the disformal Schwarzschild spacetime in DHOST theories on the orbit of the S2 star orbiting Sgr A* in the central region of our galaxy. With the publicly available astrometric and spectroscopic data for the S2 star, we perform a MCMC Bayesian analysis to make a constraint on the disformal parameter. Our results show that in the $1$-$\sigma$ range the disformal parameter is constrained to $\alpha=0.0012^{+0.0029}_{-0.0029}$. This means that the constraint of $\alpha$ from the S2 orbit is more precise than that from quasi-periodic oscillations \cite{2021arXiv210311788C} where the disformal parameter is limited in $\alpha=-0.01_{-0.012}^{0.011}$ at confidence level at $68.3\%$. Moreover, we find that the case of $\alpha=0$  still lies in the range of $1$-$\sigma$, which means that general relativity remains to be consistent with the observation data of the S2 orbit around the black hole Sgr A* in our galaxy.

    \section{\bf Acknowledgments}

    We would like to thank Prof. Tao Zhu for his useful discussions and helps.  This work was supported by the National Natural Science Foundation of China under Grant No.12275078, 11875026, 12035005,  2020YFC2201400, and the innovative research group of Hunan Province under Grant No. 2024JJ1006.

    \bibliography{s2fitting20240704}

\end{document}